\documentclass[aps,twocolumn,prd,showpacs,nofootinbib,preprintnumbers]{revtex4-1}
\usepackage{bm}
\usepackage{latexsym}
\usepackage{natbib}
\usepackage{url}
\usepackage{dcolumn}
\usepackage{color}
\usepackage{amsfonts,amssymb,amsmath}
\usepackage{graphicx,epsfig}
\usepackage{psfrag}
\usepackage{subfigure}
\usepackage{natbib}

\begin{document}

\title{Extended galactic rotational velocity profiles in $f(R)$ gravity background}

\author{Vipin Kumar Sharma}
\email[]{vipinastrophysics@gmail.com}
\author{Bal Krishna Yadav}
\email[]{balkrishnalko@gmail.com}
\author{Murli Manohar Verma}
\email[]{sunilmmv@yahoo.com}

\affiliation{Department of Physics, University of Lucknow, Lucknow 226 007, India}
\date{\today}
\begin{abstract}
An attempt has been made to explore the galactic dynamics via the rotational velocity beyond the Einstein's geometric theory of gravity. It is inspired from the geometric relation obtained in the power law $f(R)$ gravity model in vacuum. We analyse the action with a small positive deviation from the Einstein-Hilbert action (taking $R$ as $f(R)\propto R^{1+\delta}$)  at the galactic scales for the explanation of cosmological dark matter problem and obtain the contribution of dynamical $f(R)$ background geometry in accelerating the test mass. In the weak field limits, we obtain the effective acceleration of the test mass due to a massive spherically symmetric source in $f(R)$ background and develop an equation for the rotational velocity. We test the viability of the model by tracing the motion of test mass outside the typical galactic visible boundaries without considering any dark matter halo profile. We obtain a nice agreement in the outer regions (up to few tens of kpc beyond the visible boundary) of the typical galaxy by using the known galaxy data.\\  We further explore the galactic dynamics for a galaxy NGC 1052 of which the dark matter deficient galaxies, i.e., DF2 and DF4 are a part (satellite galaxies) and discuss plots of the dynamical feature of rotation curves in $f(R)$ background for the model parameter $\delta<<1$ and interpret the results for its satellite galaxies.
\end{abstract}
\maketitle
\section{\label{1}Introduction}
The theoretical modelling of the observed strange behaviour of the galactic rotation curves \cite{b1,b010,b2,b3}, gravitational lensing phenomenon (in the case of special cluster (1E 0657-558) \cite{b4,b5}, the Supernovae Ia observation \cite{b6,b7,b8}, etc., in the standard General Relativity (GR) theory  of gravity motivates one to include some dark components about 95{\%} of the total energy content of the universe. This dark sector explains well the galactic dynamics, the lensing effect and also the present accelerated expansion of the universe. These dark components get revealed in the two different flavours at different cosmological scales (galactic and extra-galactic): (i) dark matter ($\approx$27{\%}) with zero pressure which is further categorized into baryonic (like Massive Astronomical Compact Halo Objects-MACHOs) or non-baryonic (like non-standard elementary particles -- axions), (ii) dark energy ($\approx$68{\%}) with negative pressure, following the Planck's data 2018 \cite{b9}.\\ Though, still today we are not able to detect the dark matter experimentally through a number of different particle physics experiments, viz., LHC, IceCube, XENON-100, DAMA/LIBRA, CDMS II, etc. However, we know that it may exist because of its observable cosmological consequences.\\
Alternatively, it is possible that it may be the effect of modified gravity. However the concentration of dark component is different at different scales as suggested by the observations. At the scales of the solar system, visible matter is much greater than the invisible (dark) matter as predicted by the perihelion of a planetary orbit and the Keplerian curve in the weak field limit in GR \cite{b10,b11}. But, as we go beyond these scales i.e., at the galactic scales, the dark component reveals its effect on the rotational velocity curve because $H_{\alpha}$ and radio HI observations of typical galaxies suggest that the rotational curves must remain flat out to several tens of kiloparsecs. It is interpreted to indicate the existence of some invisible and strange component in the universe that must influence other objects including photons through gravitational interaction \cite{b12}.\\
However, since we could not yet detect its nature experimentally,  we have a good reason to work with the modified gravity theory. Also, in the wake of the recent observation of new galaxies \cite{b13} which lack this dark component (dark matter) we feel strongly motivated  to explore this theory.\\ Hence, we modify the Einstein-Hilbert action in order to incorporate such extra and unknown components without assuming any change in the energy-momentum tensor of the standard matter $T^{(m)}_{\mu\nu}$. This modification will generally appear as a scalar field (scalaron) in the Einstein frame action \cite{b14,b15,b105}. Thus, one can generally explain the effects of cosmological dark matter and cosmological dark energy by invoking a scalar field instead of the conventional dark sector in the cosmological models \cite{b106}. Such model turns out to be scalar-tensor model of gravity \cite{b16}. The scalar field so introduced by the modification of spacetime curvature in an Einstein frame possesses a characteristic like that of a chameleon \cite{b17,b18}. Therefore, it may mimic the role of dark matter as well as dark energy at different cosmological scales. Although there exist several models of modified gravity to explain the dark sector issues \cite{b19}, but we do not consider them here and, instead, focus on the simplest modification in order to tackle the dark matter issue at the galactic scales that can be well explained by the scalar field appeared due to the modification of gravity \cite{b20,b208}. Also, as an important manifestation of the scalar field, an arrow of time has been shown from the motion of the scalar field  at large scales \cite{b21,b201,b209}.  Thus, we work with the specific $f(R)$ model which has a small positive deviation parametrized by $\delta$ from the standard Einstein-Hilbert action at the galactic scales by taking the Ricci scalar function of the form $f(R)=\frac{R^{1+\delta}}{R_{c}^{\delta}}$ (${R_{c}}$ is a weight constant in the $f(R)$ background having the dimension of Ricci scalar). The modified gravity theory is actually inherited with an extra degree of freedom that may explain the galactic cosmological dark matter issue. In general, such theory is useful for developing a form of potential which differs from the Newtonian one but recovers it approximately in the weak field limits. \\ Thus, the key idea in the present paper is to modify the spacetime geometry according to the form of the power law $f(R)$ gravity model in order to explain the galactic signature of the cosmological dark matter. We obtain an effective acceleration equation in the presence of a massive spherically-symmetric source of mass $M$ for a test mass (or test star). Next, we obtain the equation for the rotational velocity of the test mass in dynamical $f(R)$ background and then trace the motion of the test mass outside the mass distribution source $M$ by considering it as a constant. In general, we study the simple situation for a typical galactic mass (the masses of the visible portions of the typical galaxies are estimated to lie in the range over few $ 10^9 - 10^{12}$ solar mass) with the typical velocity profile in the range $200-300$ km s$^{-1}$ \cite{b23,b24,b25}. Thus, for tracing the motion of the test mass outside the galaxy according to the behaviour of rotational velocity, we tune the $f(R)$ background galactic length scale constant $r_0$ for some fixed smaller value of the model parameter, $\delta$. This enables us to obtain the desired rotational velocity profile outside the visible boundary of the typical galaxy in dynamical $f(R)$ background, i.e., $f(R)\neq R$.\\ Hence, accordingly our work is organized as follows. In Section II, the dynamics of $f(R)$ dark matter model inspired by the Ricci scalar curvature $(R)$ is discussed. In Section III, the potential of $f(R)$ background is discussed and the rotational velocity equation of a test mass is calculated. We explore the galactic dynamics beyond the visible boundaries by using the known typical galactic specifications for the $f(R)$ rotational velocity equation in Section IV. Further, we attempt to explore the galactic dynamics for NGC 1052 galaxy in $f(R)$ background via rotational curve and interpret the results for its satellite galaxies. We conclude and discuss our work in Section V. \\ Throughout the paper, we use the signature of the spacetime metric as ($-$,+,+,+), the indices $\mu,\nu$=0,1,2,3 and c=$\hbar$=1.
\section{\label{2}Curvature scalar inspired $f(R)$ dynamics}
The main motivation to work with the fourth order gravity is that it may fill the gap of the dark sector components without actually demanding their need in strange form. Here, we assume that the spacetime is homogeneous, isotropic and spatially flat. This is given by the Friedmann-Lemaitre-Robertson-Walker (FLRW) spacetime metric as\begin{eqnarray}
ds^{2}= -dt^{2} + a^2(t)[dr^2 + r^2 (d\theta^{2} + \sin^{2}\theta d\phi^{2})] \label{a1},\end{eqnarray}
where $a(t)$ is the time dependent cosmological scale factor and ($r$,$\theta$,$\phi$) are the usual spherical co-ordinates. The $f(R)$ dynamics in such spacetime is governed by the 4-dimensional modified gravity action with standard matter as
\begin{eqnarray}
\mathcal{A}= \int d^{4}x \sqrt{-g} \left[\frac{1}{16\pi G}f(R) + L_{m}(g_{\mu\nu}, \Psi_{m})\right] \label{a2},\end{eqnarray}
where $L_m$ is the Lagrangian of the standard matter component with the matter field $\Psi_m $, $g$ is the determinant of the metric tensor $g_{\mu\nu}$ and $G$ is the Newtonian gravitational constant.
Now, the modified gravity field equations of motion is obtained by varying the action integral (2) w.r.t. $g_{\mu\nu}$
\begin{eqnarray}
F(R) R_{\mu\nu} -\frac{f(R)g_{\mu\nu}}{2} -{\nabla_\mu} {\nabla_\nu} F(R)+\nonumber\\ g_{\mu\nu}\Box F(R)= 8\pi G T^{(m)}_{\mu\nu} \label{a3},\end{eqnarray}
where $F(R)$ is the first derivative of $f(R)$ w.r.t. the Ricci scalar $R$, $R_{\mu\nu}$ is the Ricci curvature tensor, $\Box(\equiv {\nabla^\mu} {\nabla_\mu})$ is the covariant Laplacian, $T^{(m)}_{\mu\nu}$ is the energy-momentum tensor of matter and ${\nabla_\mu}$ is the covariant derivative associated with the Levi-Civita connection of the metric. We can also rewrite the equation (\ref{a3}) in the form of modified Einstein tensor equations as
\begin{eqnarray}
G_{\mu\nu}=\frac{8\pi G}{F(R)}\left[ T^{(c)}_{\mu\nu}+ T^{(m)}_{\mu\nu} \right] \label{a4},\end{eqnarray}
where \begin{eqnarray}
T^{(c)}_{\mu\nu}=\frac{1}{8\pi G}[\frac{1}{2} g_{\mu\nu}f(R)-\frac{R}{2} g_{\mu\nu} F(R)+\nonumber\\ \nabla_\mu\nabla_\nu F(R)- g_{\mu\nu} \Box F(R)] \label{a5},\end{eqnarray}
is the energy-momentum tensor of the spacetime curvature. Here, it is to be noted that the Ricci scalar is dynamical if $f(R) \neq R$, otherwise the theory gets reduced to the standard GR. Also, the constant function, $f(R)$ added or subtracted from the Ricci scalar of the Einstein-Hilbert action acts like the Einstein's cosmological constant. Thus, the dynamical form of $f(R)$ introduces a new degree of freedom  called scalaron which can explain various cosmological phenomena. \\ Now, from  equation (\ref{a1}), we can obtain an equation for the Ricci scalar
\begin{eqnarray}
R=6[2{H^2}+ \dot{H}] \label{a6},\end{eqnarray}
where ${H}$ is the Hubble expansion parameter and over dot represents the time derivative.
Also, from the field equations of $f(R)$ for the vacuum case we find
\begin{eqnarray}
3{F(R)}{H^2}=\frac{1}{2}\left[{F(R)R-f(R)}\right]- 3{H \dot F(R)} \label{a7}.\end{eqnarray}
Here, we address the problem of dark matter at the galactic scales by considering the modification of gravity having the modification of the form
\begin{eqnarray}
f(R)=\frac{R^{1+\delta}}{R_{c}^{\delta}} \label{a8},\end{eqnarray}
where $R_c$ is the weight constant and has the dimension of Ricci scalar and $\delta$ is a dimensionless quantity. Such model can explain the physical effect of dark matter on the rotation velocity profile of galaxies in $f(R)$ background.\\
We now proceed to obtain the geometry of the spacetime for the proposed model in vacuum. Therefore, on making use of equation (\ref{a6}), we express the time derivative ($\frac{d}{dt}$) as
\begin{eqnarray}
\frac{d}{dt}\equiv\frac{dH}{dt}\frac{d}{dH}=\dot{H}\frac{d}{dH}=\left(\frac{R}{6}-2{H^2}\right)\frac{d}{dH} \label{a9}.\end{eqnarray}
Next we use equation (8) and (9) in equation (7) and obtain
\begin{eqnarray}
\frac{(R-12H^2)}{(\delta-1)}\left[H\frac{dR}{dH}\delta(1+\delta)-R\delta\right]=6RH^2\label{a10}.\end{eqnarray}
On using equation (6) in equation (10), we get
\begin{eqnarray}
 6\delta\dot H \left[H \frac{dR}{dH}(1+\delta)-R \right]- \dfrac{R^2 (\delta-1)}{2} + \nonumber\\3 R \dot{H} (\delta-1) = 0 \label{a11},\end{eqnarray}
which can be re-written as
 \begin{eqnarray}
\frac{\delta (1+\delta)}{(1-\delta)}\frac{\dot R}{RH}+\left[\frac{\delta}{(1-\delta)}\left(-\frac{\dot H}{H^2}\right) + 1\right] = 0 \label{aa12}.\end{eqnarray}
To show the attractive nature at the galactic scales or deceleration, we assume $-\frac{\dot H}{H^2}>>1$ in our analysis of cosmological dark matter explanation with a much small deviation from the Einstein-Hilbert lagrangian at the galactic scales. Now, equation (12), under our assumption becomes
\begin{eqnarray}
 \frac{\delta (1+\delta)}{(1-\delta)}\frac{\dot R}{RH}+\frac{\delta}{(1-\delta)}\left(-\frac{\dot H}{H^2}\right)= 0 \label{aab12}.\end{eqnarray}
 The solution of equation (13) gives the relation between $R$ and $H$ as
\begin{eqnarray}
R=R_0 \left( \frac{H}{H_0}\right) ^{\frac{1}{1+\delta}} \label{a12},\end{eqnarray}
where $R_0$ and $H_0$ are constants and represents the present value of the Ricci scalar curvature and Hubble expansion parameter respectively.

\section{\label{3}Calculation of Galactic Rotational velocity in $f(R)$ background}

\begin{figure}[h]
\centering
\includegraphics[width=0.45 \textwidth,origin=c,angle=0]{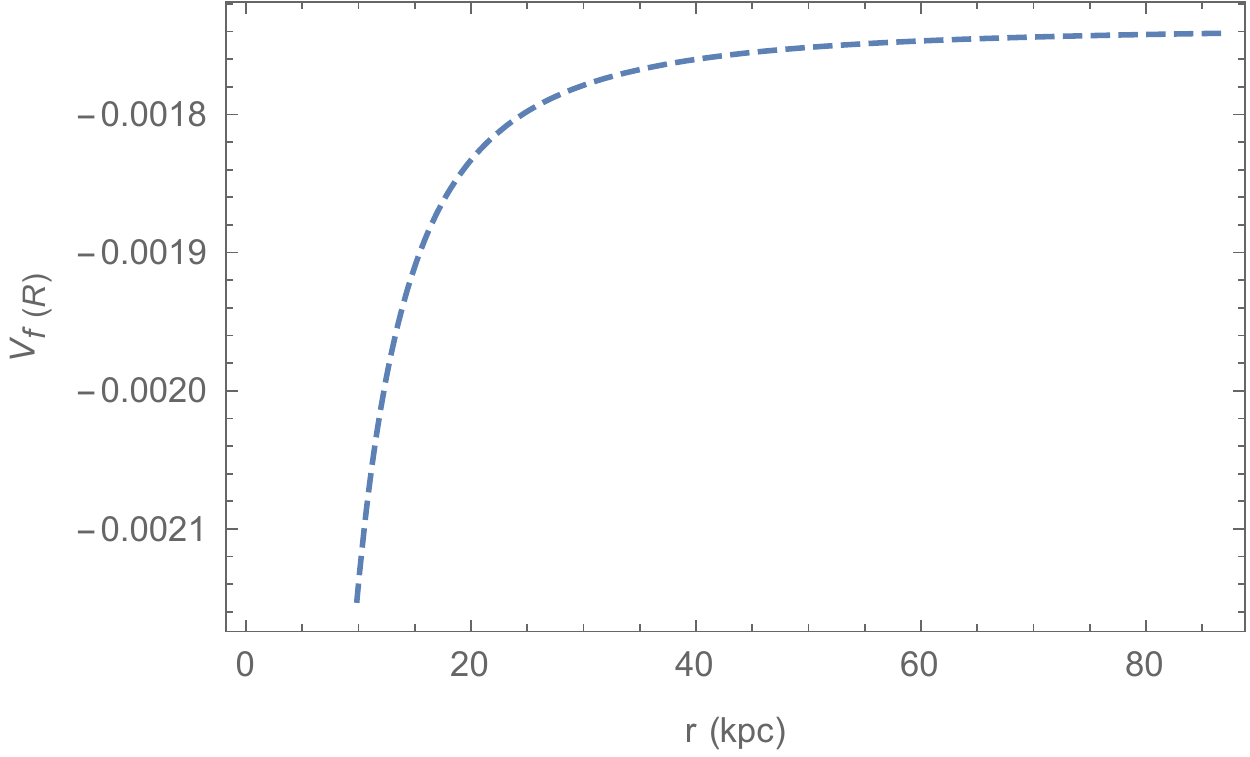}
\caption {$f(R)$ background potential vs. distance plot.  The plot is obtained by normalizing the values of all the constants appearing in equation (19) to unity with $\delta<<1$.}\label{f1}
\end{figure}

To obtain the rotational velocity in the  modified gravity theory, we start with equation (\ref{a9}) and calculate the behaviour of the scale factor $a(t)$ for the proposed form of $f(R)$ model. Thus, with equation (\ref{a12}), we get

\begin{eqnarray}
\int\frac{da}{a}=\int{\frac{H dH}{\frac{R}{6}-2 H^2}}\label{aaa13},\end{eqnarray}
which gives,
\begin{eqnarray}
{a(t)}= {a_0}\left[ \frac{1-\chi}{(\frac{H}{H_0})^{\frac{1+2\delta}{1+\delta}}-\chi}\right] ^{\frac{1+\delta}{2+4\delta}} \label{a13},\end{eqnarray}
where $\chi=\frac{R_0}{12 {H_0}^2}$ and ${a_0}$ is the present value of the cosmological scale factor.
Now, equation (\ref{a13}) can be recast as
\begin{eqnarray}
H=\frac{\dot a}{a}={H_0}\left[ \chi-(\chi-1)\left( \frac{a}{a_{0}}\right)^{-\left( \frac{2+4\delta}{1+\delta}\right)}\right]^{ \frac{1+\delta}{1+2\delta}}  \label{a14}. \end{eqnarray}
We can easily obtain the contribution of dynamical $f(R)$ background geometry in the acceleration of the test mass. So, from the equation (17), we find the acceleration equation of the test mass as
\begin{eqnarray}
\ddot{a}=\frac{{H_0}^2}{a^3}\left[ \chi {a}^\frac{2+4\delta}{1+\delta}-(\chi -1){a_0}^\frac{2+4\delta}{1+\delta} \right] ^{ \frac{1}{1+2\delta}\ }\times \nonumber\\ \left[ \chi{a}^\frac{2+4\delta}{1+\delta}+(\chi -1){a_0}^\frac{2+4\delta}{1+\delta}\right]  \label{a15}.\end{eqnarray}
Since $a_0>a$ always, so the acceleration of the test mass in $f(R)=\frac{R^{(1+\delta)}}{R_{c}^{\delta}}$ type models is negative for $\chi\approx\frac{1}{2}$ (obtained by using the current standard model of cosmology i.e., $\Lambda$CDM model, which estimates the value of $R_0 \approx 6 H_0^2$).
Hence, the potential of the $f(R)$ background having the test mass in the absence of matter will be given by using the basic kinematic definition ${V_{f(R)}}=- \int{\ddot{a}\ da}$. Thus, we have from equation (18), the potential function as (in the Planckian system of units)
\begin{eqnarray}
{V_{f(R)}}=-\frac {{H_0}^2}{2r^2}\left[ \chi {r}^{\frac{2+4\delta}{1+\delta}}-(\chi-1){r_0}^{\frac{2+4\delta}{1+\delta}}\right] ^{\frac{2(1+\delta)}{1+2\delta}} \label{a17},\end{eqnarray}
where ${r_0}$ is the $f(R)$ background galactic length scale parameter measured in kpc. The behaviour of the $f(R)$ background potential w.r.t. distance is shown in Fig. \ref{f1}.\\
It is interesting to ascertain from the modified potential a constant rotational velocity profile for the test mass \cite{b205}.
\\ Therefore, we consider a massive spherically symmetric source of mass $M$ in such $f(R)$ background. In such case, we now have the effective potential due to mass $M$ and that due to $f(R)$ background in the weak field approximation as
\begin{eqnarray}
{V_{eff}}= V_{source}+ {V_{f(R)}} \label{a18},\end{eqnarray}
where, $V_{source}$ is the simple Newtonian potential due to mass $M$ (=$-\frac{GM}{r}$) and  $V_{f(R)}$ is the geometric background potential given by equation (19). The Newtonian profile is approximated for the test mass motion closer to mass concentration.
Therefore, the effective acceleration of the test mass will be simply written by using equation (18) and the usual Newtonian acceleration in the weak field limit as
\begin{eqnarray}
{\ddot{r}}_{eff}=\frac{-GM}{r^2}-\left( \chi \right) ^\frac{2+2\delta}{1+2\delta} {H_0}^2 r \left[ 1+\frac{(\chi-1)}{\chi}\left( \frac{r_0}{r}\right) ^{2(\frac{1+2\delta}{1+\delta})}\right]\times \nonumber\\ \left[ 1-\frac{(\chi-1)}{\chi}\left(\frac{r_0}{r}\right) ^{2(\frac{1+2\delta}{1+\delta})}\right] ^{\frac{1}{1+2\delta}}\label{a19}.\end{eqnarray}
The effective potential for a massive spherically symmetric source of our proposed $f(R)$ model will solely depend on the radial distance $r$.  Using the basic kinematic definition, we have the rotational velocity of the test mass at a distance $r$ from the source in $f(R)$ background
\begin{eqnarray}
{v}\simeq\sqrt{\mid{\ddot{r}}_{eff}\  r \mid} \label{a20}.\end{eqnarray}
Thus from equations (\ref{a19}) and (\ref{a20}), we have
\begin{eqnarray}
{v}^2\simeq{\frac{GM}{r}+\left(\chi \right)^{(\frac{2+2\delta}{1+2\delta})}{H_0}^2\ r^2\ \left[ 1+\frac{(\chi-1)}{\chi}\left( \frac{r_0}{r}\right)^{2(\frac{1+2\delta}{1+\delta})}\right]} \times \nonumber\\ {\left[ 1-\frac{(\chi-1)}{\chi}\left( \frac{r_0}{r}\right)^{2(\frac{1+2\delta}{1+\delta})}\right] ^{\frac{1}{1+2\delta}}}\label{a21}, \end{eqnarray}
where the current value of $H_0$ (from the Planck data), $H_0 = 67.4\pm0.5$ km s$^{-1}$ Mpc$^{-1}$ \cite{b9}, the value of the Newtonian gravitational constant, $ G = 4.3\times 10^{-6}$ kpc km$^{2}$ sec$^{-2}$ $M$ $_{\odot}^{-1}$ \cite{c1}  and  $\chi(=\frac{R_0}{12 H_0^2})\approx\frac{1}{2}$ .\\
The first term of equation (23) predicts the Keplerian curve for the constant mass whereas the second term due to the $f(R)$ background contribution. As we go away from the mass concentration $M$, the anomaly in the rotational velocity can be explained without any dark matter through the mild modification of gravity. The behaviour of the $f(R)$ rotational velocity curve for different values of $\delta$ is plotted in Fig. 2 and Fig. 3 .
\begin{figure}[h]
\centering  \begin{center} \end{center}
\includegraphics[width=0.44 \textwidth,origin=c,angle=0]{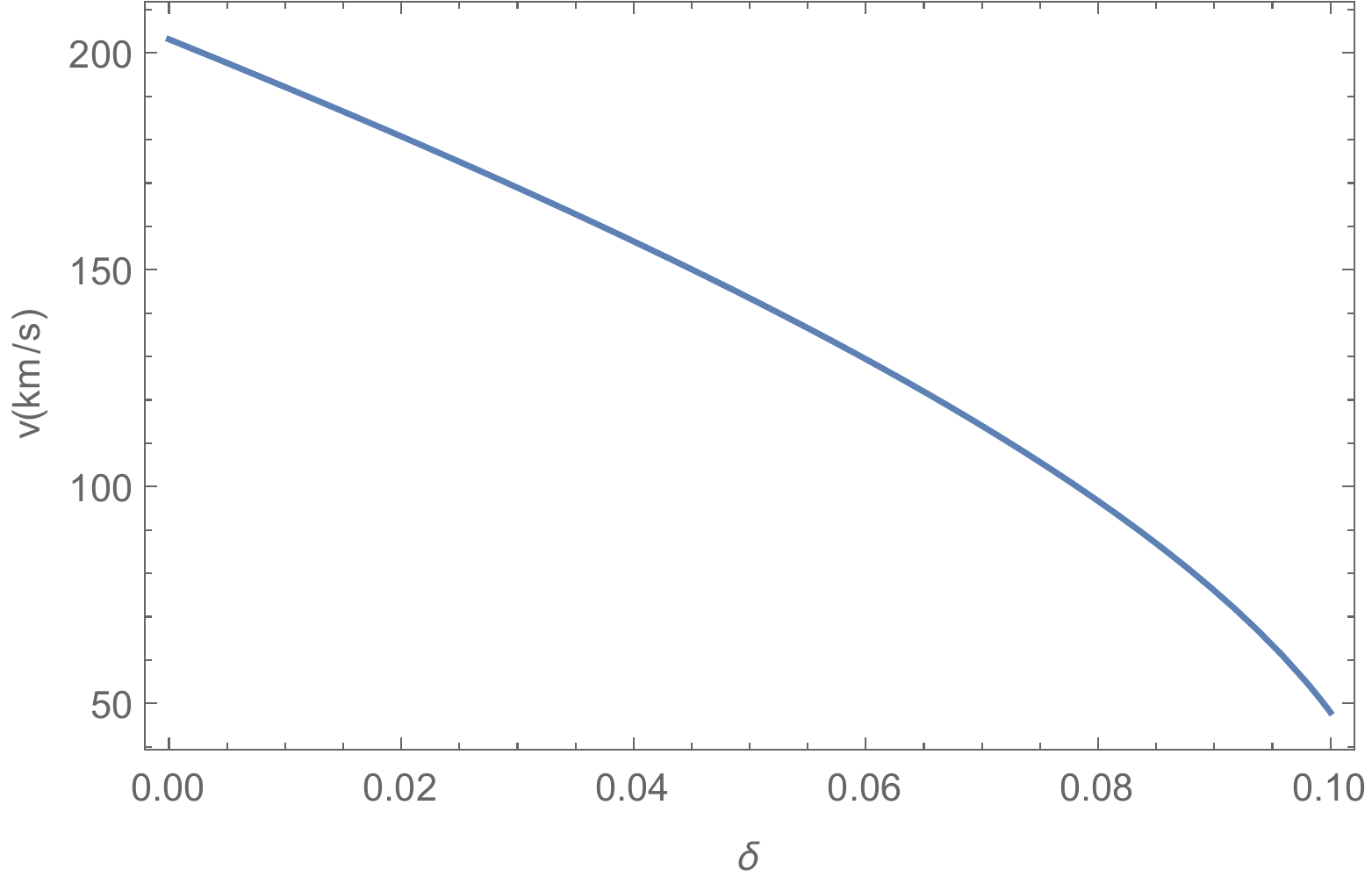}
\caption{\label{fig:p2} The theoretical effective $f(R)$ rotation velocity curve for a typical massive galaxy shows the variation w.r.t the  parameter $\delta<1$ with $\frac{r_0}{r}>1$.}\label{f2}
\end{figure}

\begin{figure}[h]
\centering  \begin{center} \end{center}
\includegraphics[width=0.44 \textwidth,origin=c,angle=0]{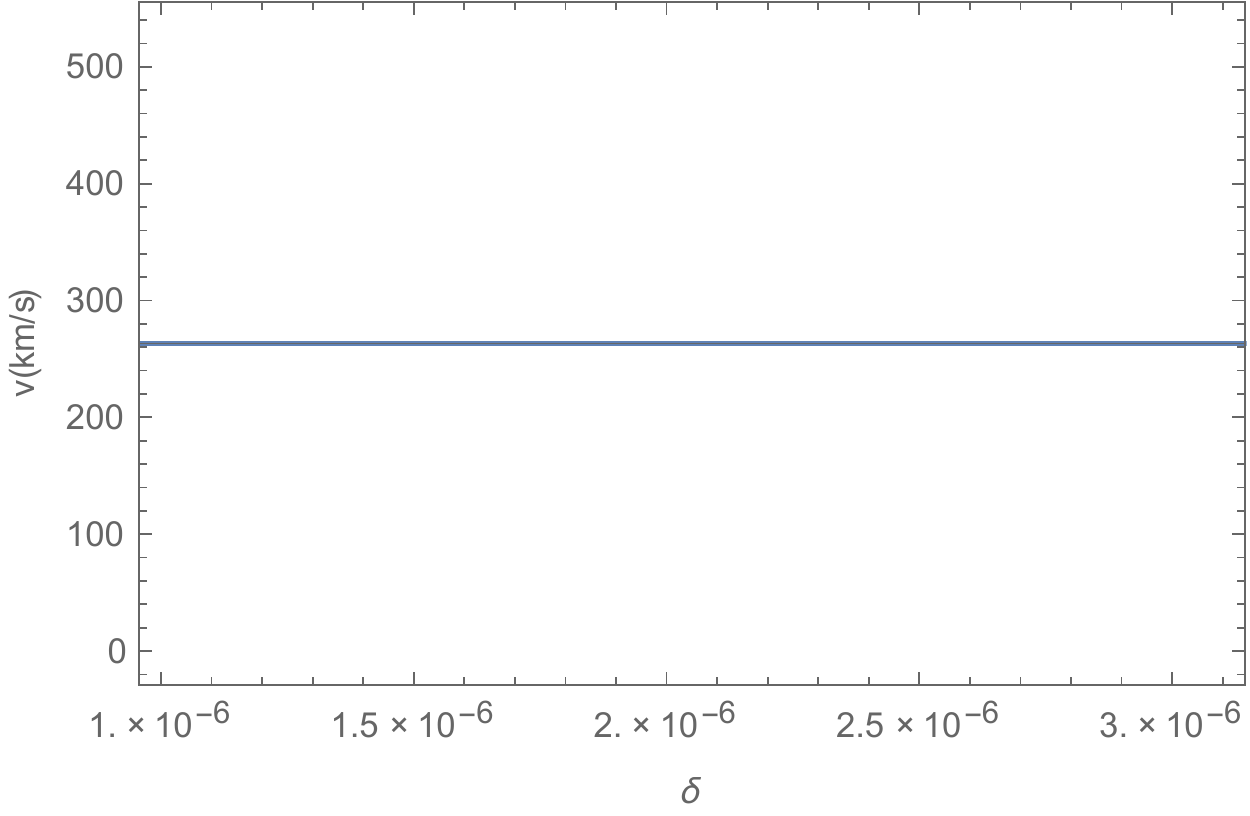}
\caption{\label{fig:p3} The theoretical effective $f(R)$ rotation velocity curve w.r.t the model parameter $\delta<<1$ with $\frac{r_0}{r}>1$. It shows that the rotational velocity attains a constant value for different values of the parameter $\delta$. The smaller value of $\delta$ seems to be consistent for the explanation of the effects of dark matter on the rotational velocity of the test mass in the outer regions of galaxy.}\label{f3}
\end{figure}
Fig. 2 shows that $f(R)$ rotational velocity decreases for slightly higher values of $\delta$ while in Fig. 3 it remains constant for $\delta\approx10^{-6}$. Hence, in contrast to both figures, the choice of smaller value of  model parameter $\delta$ is preferred. The rotational velocity is unaffected for different smaller value of the model parameter $\delta$. Thus it is interesting to explore the galactic dynamics for the smaller value of $\delta$.
\section{\label{4} Galactic dynamics via $f(R)$ rotational velocity}
To explore the galactic dynamics beyond the visible boundaries of a typical galaxy (optical disk size as in spiral galaxy), we trace the test mass in such region according to equation (23); thus enclosing most of the visible part of the typical galaxy, one can expect the galaxy mass to be approximately constant. Therefore, the rotational velocity curve for the test mass should drop-off as the  square-root of the distance in weak, unmodified gravity. However, due to the contribution of the $f(R)$ background potential, the rotational velocity  holds a constant value for $\delta<<1$, as shown in the Fig. 3. 
\\
Further, to get the actual trace of the test mass beyond our Milky Way size in $f(R)$ background, we consider the typical visible size to be $r_{galaxy}\approx 15$ kpc (visible disk size) with the estimated mass, $M\approx (7\pm2.5)\times10^{11}M_{\bigodot}$, without explicitly considering any dark matter halo profile \cite{b24} and tune the value of $f(R)$ background galactic length scale parameter $r_0$, to obtain the observed velocity match for the $f(R)$ model parameter with ${\delta\approx10^{-6}}$. 
\\
On the basis of the plot obtained for the $f(R)$ rotational velocity in Fig. 4, we argue that beyond the typical galactic size, the rotational curve shows constant nature up to few kpc, which is a nice feature of our theoretical model. This nature arises only because of the $f(R)$ background potential. There is an initial rise observed in the rotational velocity  beyond the visible boundaries of the typical galaxy. Such bump is also observed in THINGS (The HI Nearby Galaxy Survey) rotation curve at about 15 kpc, which is considered to be the source of information of the test mass motion along the two stretched spiral arms \cite{b204}.\\ The study of certain galaxies by Rubin et al., \cite{b010} addresses the rising rotation curves at the outer radii which implies that the dark matter fraction in such galaxies increases as luminosity decreases. Since, we do not consider any dark matter profile, so it may be due to the fact that beyond the  visible boundaries and turn over radius, the physical effect of $f(R)$ background geometry becomes significant for the suitable values of the model parameter and the galactic length scale parameter and hence we get the expected result. \\
It is also important to  note that the empirical data available so far, for the typical galaxy rotational curves in the outer  region (beyond the visible end), probe the rotational velocity only up to few kpc (about 20 kpc) where the rotational velocity may reach about $220-300$ km s$^{-1}$ \cite{b24,b25}.\\
However, our work seems sufficient to study the dynamical behaviour of the rotational velocity curve beyond the typical galactic visible boundaries.
\begin{figure}[h]
\centering  \begin{center} \end{center}
\includegraphics[width=0.44 \textwidth,origin=c,angle=0]{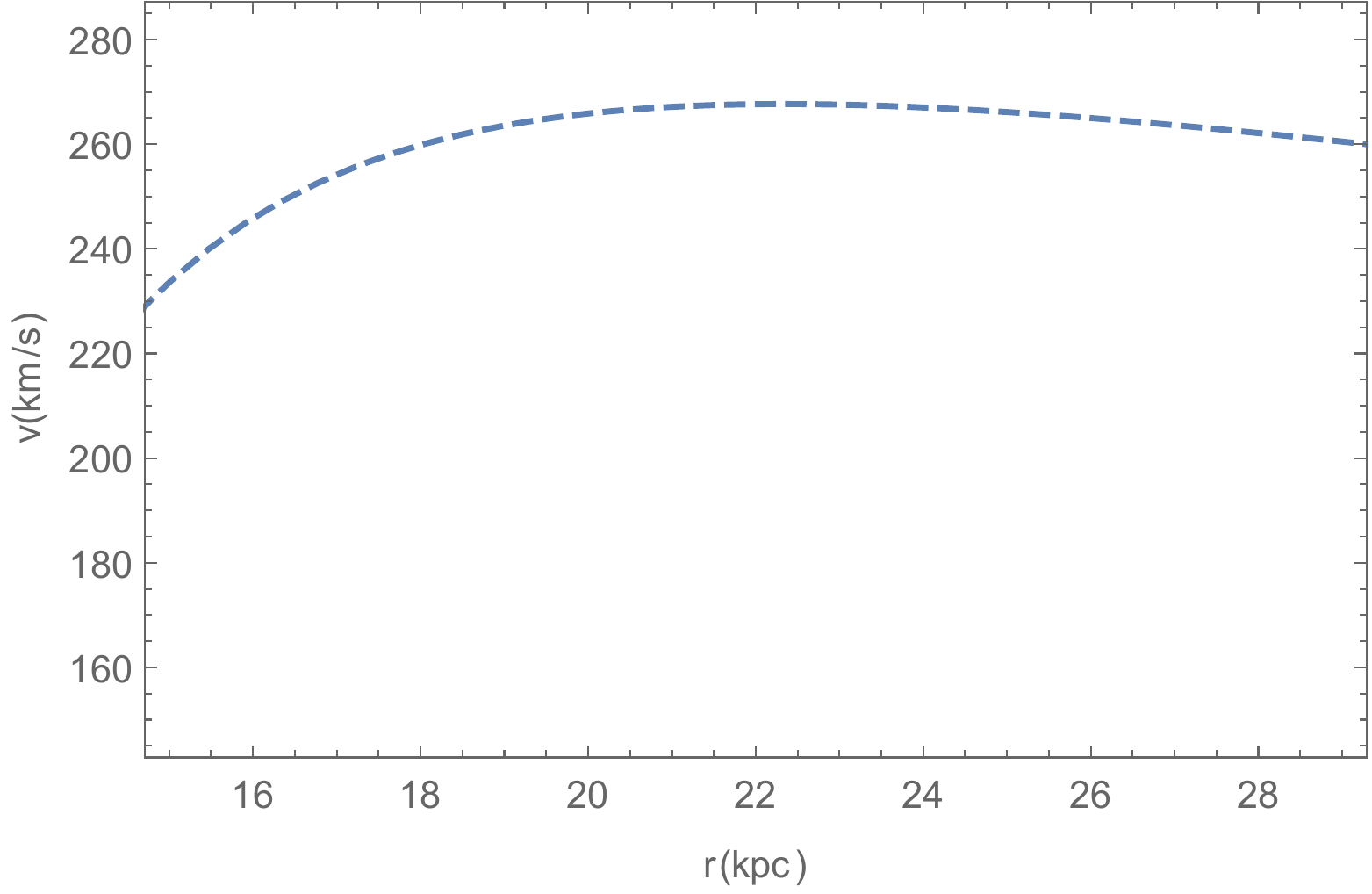}
\caption{\label{fig:p4} Theoretical galactic rotation curve external to the typical visible end of Milky Way galaxy. The curve shows the behaviour of the test mass beyond the typical massive $(7\pm2.5)\times10^{11}M_{\bigodot}$, galactic size (about 15 kpc) and is obtained for ${\delta\approx10^{-6}}$ with the galactic length scale parameter $r_0=10^{2.623}$ kpc $ \approx $ 419.280 kpc.}\label{f4}
\end{figure}
Thus, as an important diagnostic test of the proposed model for the cosmological dark matter at the galactic scales, we find a nice agreement for the profile obtained with the modified rotation velocity curves in an extended manner without any dark matter halo which matches with the observed behaviour of the rotational velocity curve beyond the visible boundaries \cite{b26}.\\
As an important consequence of this work, the galactic dynamics of the galaxies having different morphology can also be explored. Hence, we plot (Fig. 5) the approximate viable rotation velocity profile of NGC 1052 galaxy (an elliptical galaxy) which has the recently observed dark matter deficient satellite galaxies (DF2 and DF4\textemdash ultra diffuse galaxies). The size of NGC 1052 galaxy is comparable to that of the Milky Way galaxy.  From {\cite{b206}}, we take the optical and radio data for NGC1052 galaxy and plot the modified rotational velocity curve by using equation (23).
\begin{figure}[h]
\centering  \begin{center} \end{center}
\includegraphics[width=0.44 \textwidth,origin=c,angle=0]{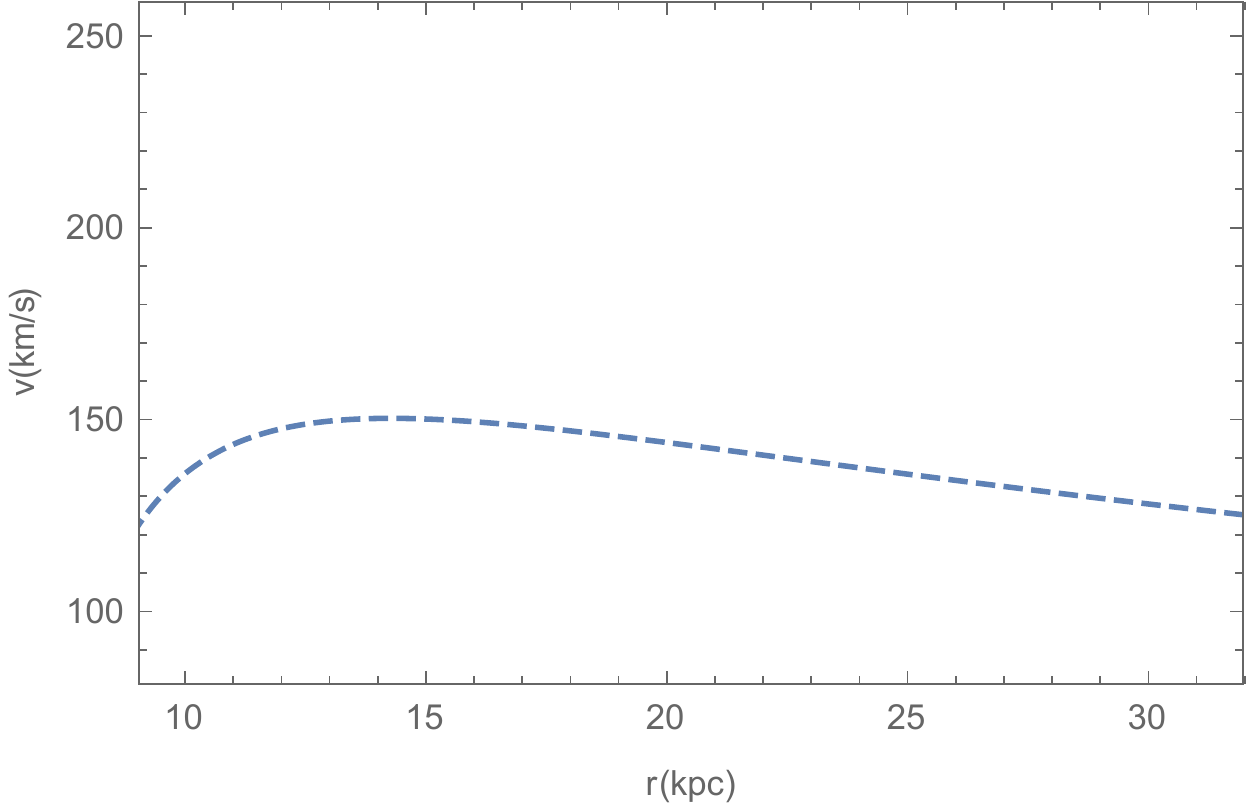}
\caption{\label{fig:p5} The rotation curve external to the typical visible end of the galaxy NGC 1052. The curve shows the behaviour in the outer region with $M=1.5\times10^{11}M_{\bigodot}$, $\delta\approx10^{-6}$ and the galactic scaling constant $r_0= 10^{2.400}$ kpc $ \approx 251.189$ kpc.}\label{f5}
\end{figure}\\
Fig. 5 shows the behaviour of the modified rotation velocity curve for NGC 1052. The curve shows the declining behaviour beyond 30 kpc and up to that scale, it shows approximately constant rotation velocity lying between $120-150$ km s$^{-1}$ for $\delta\approx10^{-6}$. The nature of our theoretical modified rotation velocity curve may differ from its actual nature (rotational velocity of the ionised gas in the stable circular orbits about 140 km s$^{-1}$ at the angular distance $>$5 arcsec {\cite{b206}} for NGC 1052 whereas the HI has the rotation velocity about 200 km s$^{-1}$ at 16 kpc from the centre of the galaxy) because of the different morphology of the galaxy. The approximately constant and extended behaviour of the outer rotation curve about up to 30 kpc of NGC 1052 in the dynamical $f(R)$ background seems to explain the recent observations. It is possible that such an extended profile may be the reason of deficiency of dark matter in its satellite galaxies. This statement can be supported via the observations of Bullet Cluster which suggest that the distribution of dark matter is uneven. However, in $f(R)$ gravity, it may be due to the anisotropic propagation of scalaron (an extra degrees of freedom) in dynamical $f(R)$ background in an Einstein frame due to the chameleon-like nature [18].  Also, beyond 30 kpc, the decline in rotation velocity is not according to the Keplerian form in $f(R)$ background. Thus, in general, our framework is  sufficient to study the dynamical feature of the modified rotational curves for different galaxies with the smaller value of model parameter ${\delta}\approx 10^{-6}$.

\section{\label{5} Summary and Conclusion}
 The explanation of the existing observed unusual results about the rotational velocity curves of a typical galaxy signifies the demand to modify the existing gravity theory at such scales without any particulate form of dark matter. It may be possible that such effects (flatness of rotation velocity curve) can be explained by  modifying the spacetime curvature. Therefore, we address the cosmological dark matter problem in a modified gravity theory by considering, $f(R)=\frac{R^{1+\delta}}{R_c^{\delta}}$ type model with small $\delta$ at the galactic scales in Jordan frame. We explain the problem by plotting the viable galactic rotation curves beyond the visible boundaries for the smaller value of $f(R)$ model parameter $\delta$. From the solution (scale factor) of the modified field equations, we generate a potential due to the $f(R)$ background for explaining the cosmological dark matter issue at the galactic scale. We consider a spherically symmetric massive source of mass $M$ in $f(R)$ background with an effective potential (Newtonian+$f(R)$) in the weak field limits. We find an expression for the modified galactic rotational velocity of a test mass. Since, the presence of $f(R)$ gravity gives an extra potential, therefore the galactic rotational curve of the test mass is different from the Keplerian curve. We have shown the variation of rotational velocity of the test mass with distance from the visible ends of the typical galaxy for the $f(R)$ model parameter ${\delta\approx10^{-6}}$ and find that the rotation velocity remains constant in the extended region up to few kpc which is greater than that of the observed typical empirical data value ($\approx$ 20 kpc). Hence, it shows the similar effect as attributed to cosmological dark matter without actually demanding the need for it. We also hope that with the much advanced instruments in near future, we can probe the edge of galaxy for optical data in a more precise way. \\
Furthermore, such framework can also be used to explain the recent observations of the ultra-diffuse satellite galaxies without any specific morphological parameter consideration. The extended rotational velocity curve of galaxy i.e., NGC 1052, much beyond its optical boundaries for $\delta<<1$ may reveal the status of dark matter in its companion galaxies (DF2 and DF4) in $f(R)$ background. The reason may be the uneven distribution of dark matter in GR background or an anisotropic propagation of scalaron in the dynamical $f(R)$ background.  We studied  such dynamical feature of the modified rotation velocity curve in $f(R)$ background. Additionally, the dynamical $f(R)$ geometric background potential enhances the strength of gravity without the introduction of matter degrees of freedom and thus explains the dynamical features at the galactic scale with the usual matter. It may also be possible in our framework to explore the size of galactic haloes for different galaxies.

\section*{Acknowledgments}
Authors thank IUCAA, Pune, for the facilities during which a part of the present work was completed under the associateship programme. VKS also thanks Varun Sahni and Swagat S. Mishra for their constant support and motivation.\\


\end{document}